\documentclass{article}
\usepackage{spconf, amsmath, graphicx, siunitx, booktabs, fixltx2e, tabularx, xcolor}
\usepackage{hyperref}

\graphicspath{{./figures/}}

\title{A robust network architecture to detect\\ normal chest X-ray radiographs
\thanks{This paper was accepted by IEEE ISBI 2020. \copyright 2020 IEEE. Personal use of this material is permitted. Permission from IEEE must be obtained for all other uses, in any current or future media, including reprinting/republishing this material for advertising or promotional purposes, creating new collective works, for resale or redistribution to servers or lists, or reuse of any copyrighted component of this work in other works.}
}
\name{\begin{tabular}{c}
Ken C. L. Wong$^{1}$, PhD; Mehdi Moradi$^{1,\textbf{*}}$, PhD; Joy Wu$^{1}$, MD MPH; Anup Pillai$^{1}$, PhD; \\
Arjun Sharma$^{1}$, MD; Yaniv Gur$^{1}$, PhD;  Hassan Ahmad$^{1}$, MD; \\
Minnekanti Sunil Chowdary$^{2}$, MD;  Chiranjeevi J$^{2}$, DNB; Kiran Kumar Reddy Polaka$^{2}$, DNB; \\
Venkateswar Wunnava$^{2}$, MSc; DC Reddy$^{2}$, PhD; Tanveer Syeda-Mahmood$^{1}$, PhD
\end{tabular}
}
\address{$^{1}$IBM Research, Almaden Research Center, San Jose, CA, USA
\\ $^{2}$The Deccan Hospital Hyderabad, India\\$^{\textbf{*}}$Corresponding author: mmoradi@us.ibm.com}
\begin{document}
%
\maketitle
\begin{abstract}

We propose a novel deep neural network architecture for normalcy detection in chest X-ray images. This architecture treats the problem as fine-grained binary classification in which the normal cases are well-defined as a class while leaving all other cases in the broad class of abnormal. It employs several components that allow generalization and prevent overfitting across demographics. The model is trained and validated on a large public dataset of frontal chest X-ray images. It is then tested independently on images from a clinical institution of differing patient demographics using a three radiologist consensus for ground truth labeling. The model provides an area under ROC curve of 0.96 when tested on 1271 images. We can automatically remove nearly a third of disease-free chest X-ray screening images from the workflow, without introducing any false negatives (100\% sensitivity to disease) thus raising the potential of expediting radiology workflows in hospitals in future.

\end{abstract}
\section{INTRODUCTION}
\label{sec:intro}

The use of artificial intelligence in radiology promises to streamline and reduce clinical workload. Common radiology exams, such as chest X-rays (CXR) are among the candidates for exploring preliminary reads. Most work in this area, however, is focused on the detection of a small number of highly common diseases from CXR images \cite{DBLP:journals/corr/abs-1901-07031,DBLP:journals/corr/abs-1711-05225,Conference:Wang:CVPR2017}, facilitated by the recent publication of major public datasets \cite{Conference:Wang:CVPR2017,Journal:Johnson:arXiv2019}. Despite the progress, two issues limit the impact of these studies. Firstly, the focus on a narrow list of diseases means that these models cannot be used as effective means of filtering. Secondly, these datasets are often obtained in one institution and the resulting models do not perform as well when tested in clinical settings with data at a different hospital or demography. In a previous work we have reported the significant drop of accuracy observed when a model trained on NIH ChestXray14 data is used on data from a different institution \cite{Conference:Madani2018}.

In this work we focus on developing a network that addresses the above issues. We focus on detecting images that show no finding of concern in CXR as a well-defined class to be distinguished from all other cases treated as abnormal.  Automatically removing the normal cases in daily CXR reads can significantly improve the radiology workflows in busy hospitals. However, for the technique to be clinically viable, it should ensure that no abnormal case is accidentally filtered out. In addition, for the time/cost savings to be significant, there should be a sufficient number of cases filtered as normals as well. Our method achieves this balance by allowing nearly a third of disease-free chest X-ray screening images to be filtered, without filtering a single abnormal case.

Our paper makes two major contributions:

\noindent - \textit{First}, we report a new deep neural network architecture for CXR classification to detect normalcy. This novel network architecture provides a high degree of accuracy and robustness by building a feature pyramid from two different pre-trained networks, and employing skip connections, dilated blocks, spatial drop-out, and group normalization to maximize the generalization capability. To prepare training data for this architecture, we have developed an NLP pipeline to isolate negative images against a comprehensive list of anatomical and lines/tubes related findings. The NLP pipeline is subsequently examined by experts through reading of evidence sentences.

\begin{table*}[t]
\caption{Summary of the types of anatomical labels used to exclude a CXR from being labeled "normal anatomically".}
    \centering
    \begin{tabular}{|l|l|}
        \hline
        Category&Labels\\
        \hline\hline
        Lungs&linear/patchy atelectasis, lobar/segmental collapse, consolidation, pulmonary edema/hazy opacity,\\
        &not otherwise specified opacity, mass/nodule, azygous fissure, cyst/bullae, hyperaeration, \\
        &increased reticular markings/ild pattern, lobectomy, vascular redistribution\\
        \hline
        Pleura&pneumothorax, pleural effusion or thickening, hydropneumothorax\\
        \hline
        Mediastinum&enlarged cardiac silhouette, superior mediastinal mass/enlargement, pneumomediastinum, \\
        &mediastinal displacement, enlarged hilum, lymph node calcification, vascular calcification, \\
        &not otherwise specified calcification, tortuous aorta\\
        \hline
        Bones&fracture, spinal degenerative changes, shoulder osteoarthritis, bone lesion, dislocation, scoliosis,\\
        &diffuse osseous irregularity, elevated humeral head, osteotomy changes\\
        \hline
        Other&hernia, elevated hemidiaphragm, subcutaneous air, sub-diaphragmatic air, bullet/foreign bodies, \\
        &contrast in the gi or gu tract, dilated bowel, other soft tissue abnormalities, post-surgical changes\\
        \hline
        Exam quality&non-diagnostic cxr\\
        \hline
    \end{tabular}
    \label{tab:opacities}
    \label{table:findings}
\end{table*}

\noindent - \textit{Second}, we report a clinical study to show the robustness of our network on new data acquired from a hospital. In this ``wild'' clinical situation, it performs with an area under receiver operating characteristic (ROC) curve of 0.96 as validated by the consensus opinion of three radiologists on the test data. We show that this network can isolate and remove 33\% of the normal images from the workflow without clinician involvement, and without missing a single disease case.

\section{METHODOLOGY}
\label{sec:method}
We first describe how we used the text reports from the MIMIC dataset to build a dataset of normal versus abnormal X-ray images. The architecture and training strategy of the deep learning model is described next. Finally, we describe our study using this model on clinical data acquired from an institution of differing demographics, and outline its ground truth generation and validation methodology.

\subsection{Labeling and data curation for training and validation}

The following is a brief description of our natural language processing (NLP) pipeline applied to create the training set from NIH and MIMIC reports. The pipeline utilizes a CXR ontology curated by our clinicians from a large corpus of CXR reports using a concept expansion tool \cite{coden2012spot} applied to a large collection of radiology reports. Abnormal terminologies from reports are lexically and semantically grouped into radiology finding concepts. Each concept is then ontologically categorized under major anatomical structures in the chest (lungs, pleura, mediastinum, bones, major airways, and other soft tissues), or medical devices (including various prosthesis, post-surgical material, support tubes and lines). Given a CXR report, the text pipeline 1) tokenizes the sentences with NLTK \cite{loper2002nltk}, 2) excludes any sentence from the history and indication sections of the report via key section phrases so only the main body of text is considered, 3) extracts finding mentions from the remaining sentences, and 4) finally performs negation and hypothetical context detection on the last relevant sentence for each finding label. Finally, clinician driven filtering rules are applied to some finding labels to increase specificity (e.g. "collapse" means "fracture" if mentioned with bones, but should mean "lobar/segmental collapse" if mentioned with lungs).

To generate the training dataset, we define a CXR to be \textit{normal} if none of the major anatomical finding labels are positively occurring,  the exam was not technically very limited (i.e. non-diagnostic), and the tubes and lines if present, were not misplaced, as documented in the corresponding radiology report. The vocabulary for tubes and lines covered all major lines and tubes found in chest X-rays ranging from central vascular lines to endotracheal tubes and gastric tubes as well as external devices such as cardiac pacemakers. The anatomical finding labels were also covered comprehensively to exclude a CXR from being labeled "normal anatomically" as listed in Table \ref{table:findings}.

\begin{figure*}[t]
    \footnotesize
    \centering
    \begin{minipage}[b]{1\linewidth}
      \centering
      \includegraphics[width=1\linewidth]{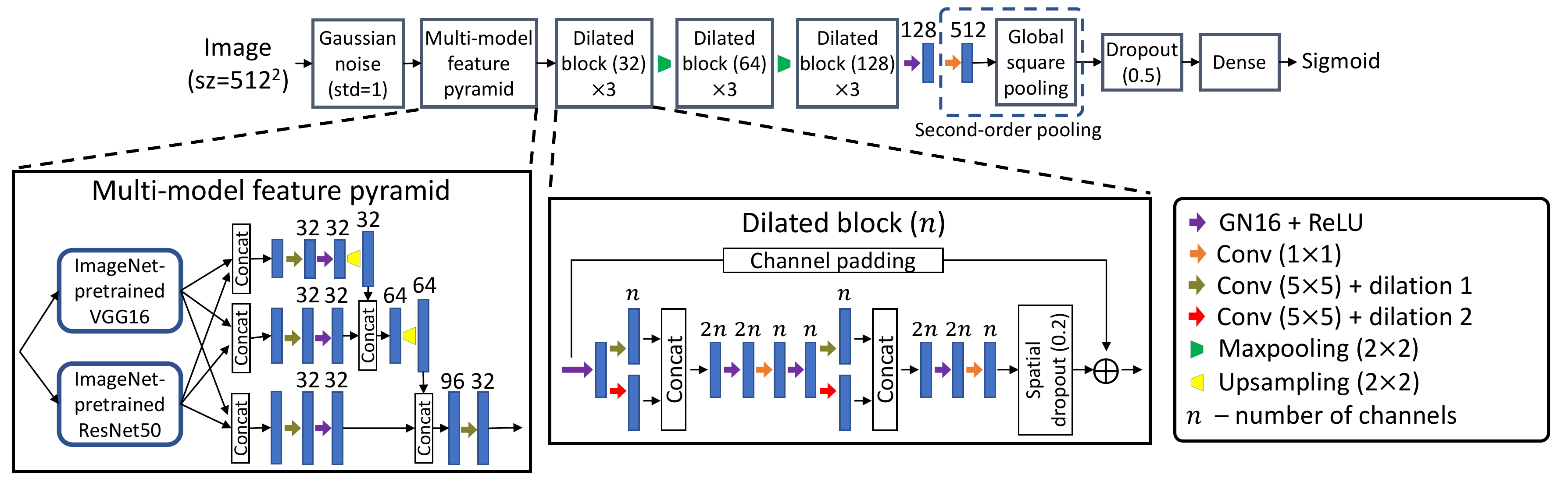} \\
    \end{minipage}
    \caption{Network architecture for VGG16+ResNet50 pyramid. Blue boxes represent feature maps. GN16 stands for group normalization with 16 groups of feature channels.}
    \label{fig:network}
\end{figure*}

This automatic labeling pipeline is checked by experts, through examining evidence sentences. Using the approach above, we analyzed nearly 220,000 reports obtained from two different sources, namely, (a) a subset of 16,000 images NIH dataset released earlier \cite{Conference:Wang:CVPR2017} relabeled by our radiologists, (b) the rest were full-fledged reports provided under a consortium agreement to us for the MIMIC-CXR dataset recently released \cite{Journal:Johnson:arXiv2019}. We retained 112,828 images for training and 16,058 images for testing. All these were frontal-view and were resized to 512$\times$512. The image intensities of DICOM images were normalized using the windowing information in the DICOM header.

\begin{table}[h]
\caption{The distribution of frontal images in the model building phase (first two lines) drawn from NIH and MIMIC datasets, and clinical study phase from partner institution (last two lines). }
\begin{tabular}{ c| c | c }
& Normal & Abnormal \\
 \hline
 \hline
 Public dataset, training & 56,406 & 56,422 \\
 Public dataset, testing  & 8,041 & 8,017 \\
 \hline
 \hline
 Clinical test data, triple consensus & 701 & 570 \\
 Clinical test data, two out of three & 948  & 801 \\
\end{tabular}
    \label{table:datacounts}
\end{table}

\subsection{Network architecture}

Compared with natural images, medical images of the same modality are visually very similar. Therefore, appropriate deep-learning techniques are required to achieve the desired fine-grained classification performance (Fig. \ref{fig:network}). The work in \cite{Conference:Nguyen:ISCAS2018} shows that concatenating different ImageNet-pretrained features from different networks can improve classification on microscopic images. Following this idea, here we combine the ImageNet-pretrained features from different models through the Feature Pyramid Network in \cite{Conference:Lin:CVPR2017}. This forms the multi-model feature pyramid which combines the features in multiple scales. The VGGNet (16 layers) \cite{Journal:Simonyan:arXiv2014} and ResNet (50 layers) \cite{Conference:He:CVPR2016} are used as the feature extractors. As natural images and CXR are in different domains, relatively low-level features are used. From the VGGNet, the feature maps with 128, 256, and 512 feature channels are used, which are concatenated with the feature maps from the ResNet of the same spatial sizes which have 256, 512, and 1024 feature channels. We also explored an alternative architecture where we replaced the ResNet50 feature extractor with DenseNet121 \cite{DBLP:journals/corr/HuangLW16a}. We report results on both architectures.

We propose the dilated blocks to learn the high-level features from the extracted ImageNet features. Each dilated block is composed of dilated convolutions for multi-scale features \cite{Journal:Yu:arXiv2015}, a skip connection of identity mapping to improve convergence \cite{Conference:He:ECCV2016}, and spatial dropout to reduce overfitting. Group normalization (16 groups) \cite{Conference:Wu:ECCV2018} whose performance is independent of the training batch size is used with ReLU. Dilated blocks with different feature channels are cascaded with maxpooling to learn more abstract features. Instead of global average pooling, second-order pooling is used, which is proven to be effective for fine-grained classification \cite{Conference:Yu:ECCV2018}. Second-order pooling maps the features to a higher-dimensional space where they can be more separable. Following \cite{Conference:Yu:ECCV2018}, the second-order pooling is implemented as a 1$\times$1 convolution followed by global square pooling.

\begin{figure*}[t]
\minipage{0.32\textwidth}
  \includegraphics[width=\linewidth]{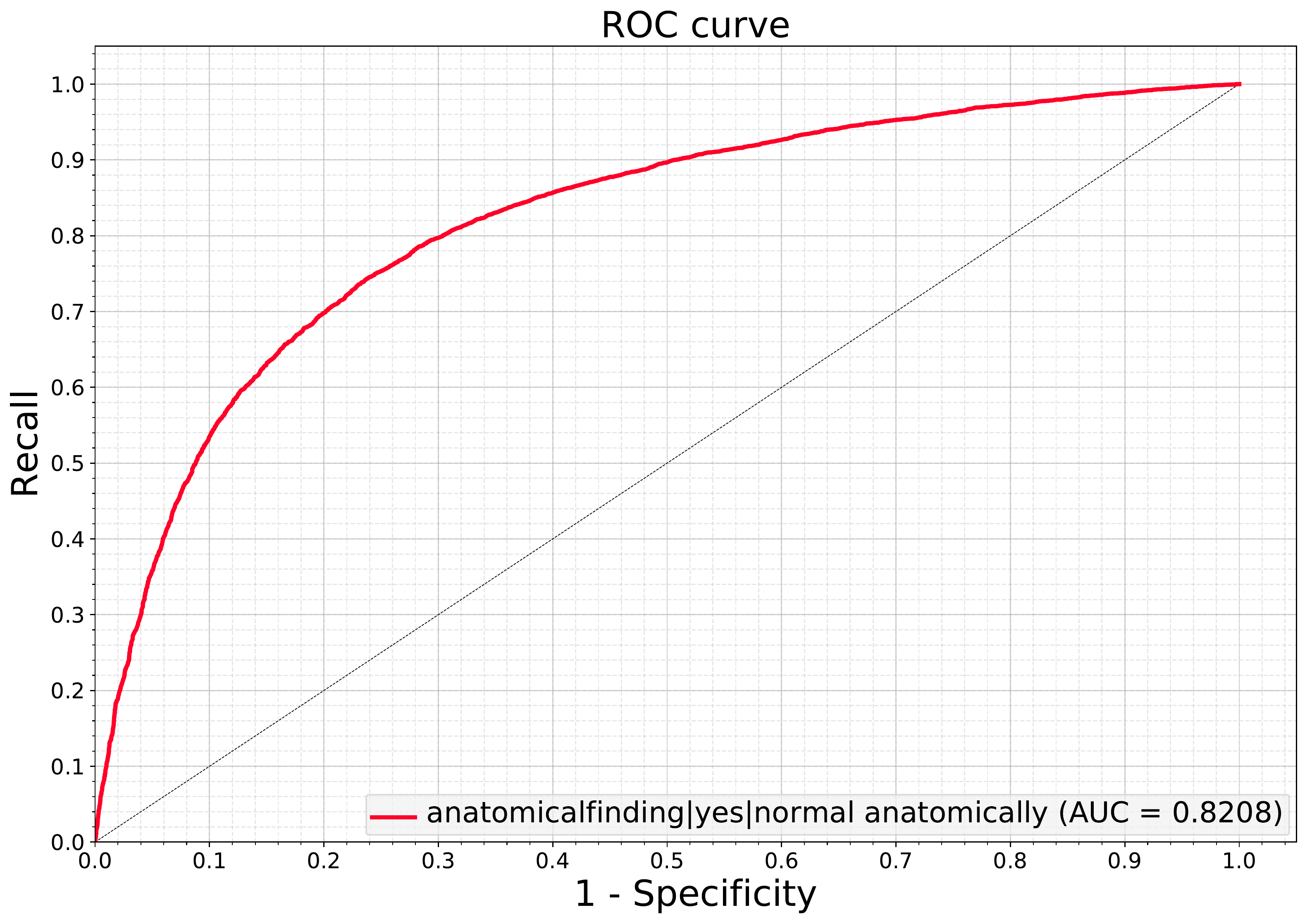}
  \centering{(a) MIMIC (AUC = 0.82)}
\endminipage\hfill
\minipage{0.32\textwidth}
  \includegraphics[width=\linewidth]{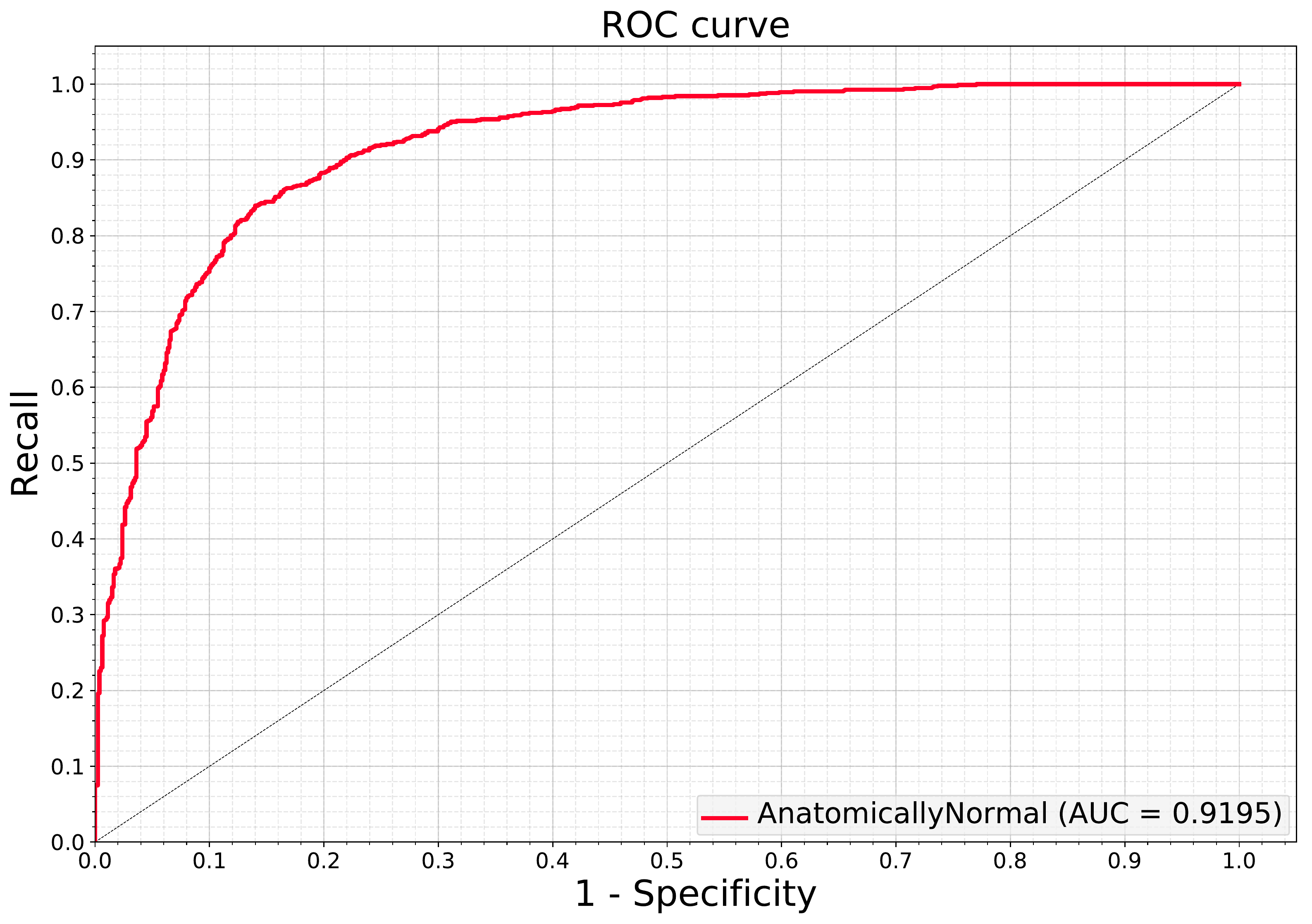}
  \centering{(b) 2/3-majority votes (AUC = 0.92)}
\endminipage\hfill
\minipage{0.32\textwidth}%
  \includegraphics[width=\linewidth]{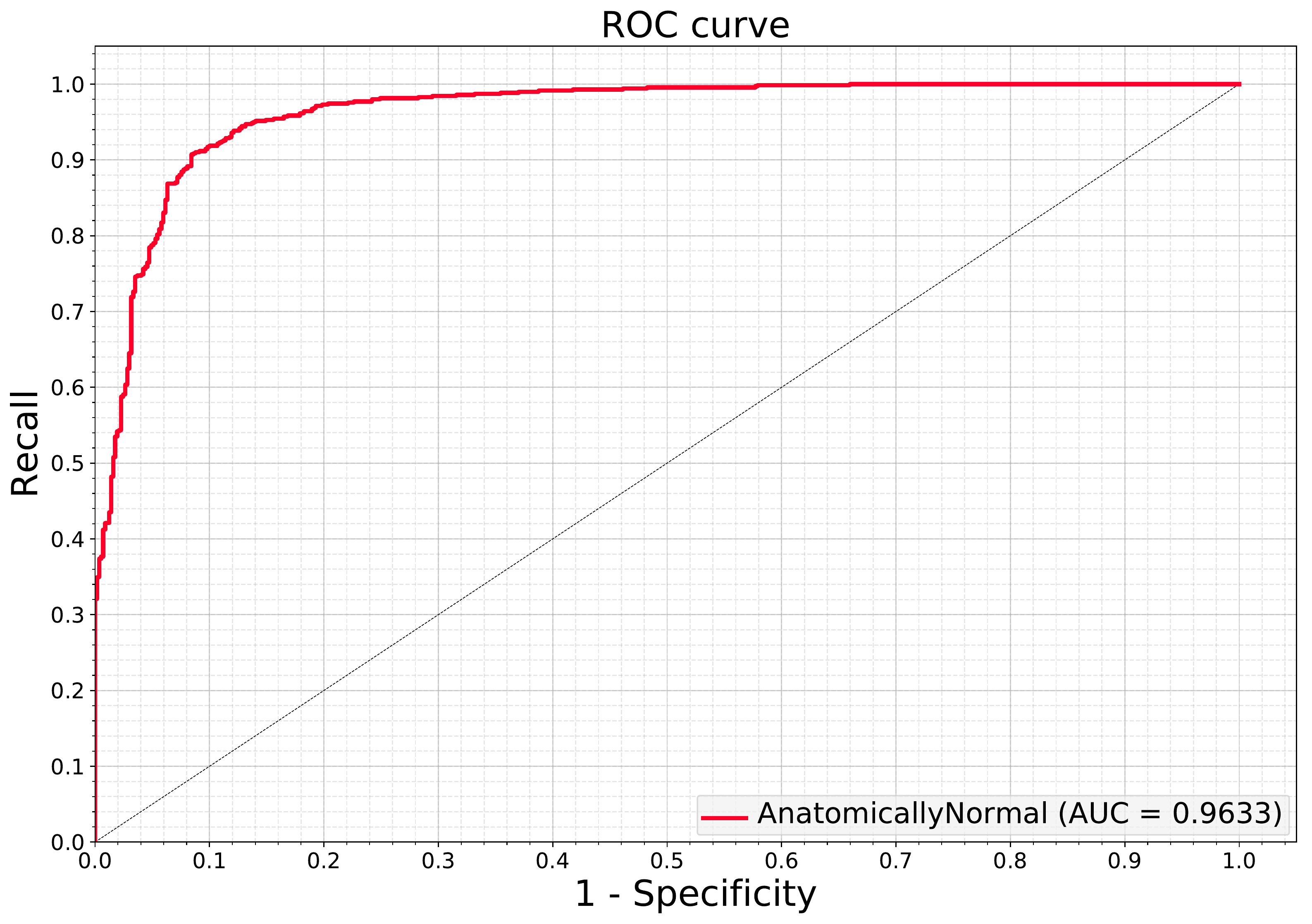}
  \centering{(c) Triple-consensus (AUC = 0.96)}
\endminipage
\caption{ROC curves of the proposed VGG16+ResNet50 pyramid architecture on different test sets. (a) The test portion of the MIMIC data formed by NLP analysis. (b) The test set from clinical practice at Deccan Hospital formed by the majority of three radiologist opinions. (c) The test set from clinical practice at Deccan Hospital formed by triple-consensus radiologist opinions.}
\label{fig:rocs_all}
\end{figure*}

\subsection{Training strategy}

Image augmentation is used to avoid over-fitting, and we limit the augmentation to rotation ($\pm$\ang{10}) and shifting ($\pm$10\%). The probability of an image to be transformed is 80\%. The optimizer Nadam is used with a learning rate of 2$\times$10$^{-6}$, a batch size of 48, and 20 epochs. The IBM Power System AC922 equipped with NVLink for enhanced host to GPU communication was used. This machine features NVIDIA Tesla V100 GPUs with 16 GB memory, and two of these GPUs were used for multi-GPU training.

\subsection{Deployment on new clinical cases: Deccan Hospital}

Through a partnership with Deccan Hospital in Hyderabad, India, with the approval of the Institutional Research Board of that institution, we performed a study of the performance of the model built on MIMIC and NIH datasets in a clinical setting. The clinical team at Deccan provided labels (normal or abnormal) for 1749 images obtained during routine clinical practice from November 2018 to April 2019. Each of these images were also de-identified and examined by two board-certified radiologists based in the United States. As Table \ref{table:datacounts} shows, in case of 1271 images, the two radiologists were in agreement with the clinical decision recorded at the hospital. This formed a triple consensus test set. We report results on both triple consensus set (1271 images) and majority vote ground truth (1749 images).

\section{RESULTS AND DISCUSSION}

\begin{table}
\caption{Area under ROC curve, and area under PR curve, obtained on test portion of the public data, using different architectures. The total number of test images is 16,058. The last row is the proposed architecture described in Methods. }
\begin{tabular}{ c | c | c }
Architecture & AUC: ROC & AUC: PR  \\
\hline
DenseNet121  & 0.590 & 0.574 \\
ResNet50 & 0.687 & 0.682 \\
\textbf{VGG16+DenseNet121 pyramid} & 0.824 & 0.817\\
\textbf{VGG16+ResNet50 pyramid} & 0.821 & 0.811
\end{tabular}
\label{table:rocs1}
\end{table}

Table \ref{table:rocs1} lists the area under ROC and precision-recall (PR) curves of four different architectures, including two baseline models without feature pyramid, on the test portion of the public datasets used for model development. It is evident that the proposed architecture,  using either VGG16+ResNet50 or VGG16+DenseNet121, provides a significant improvement in classification performance compared to baseline DenseNet121 and ResNet50 models.

Table \ref{table:rocs2} reports the area under ROC and PR curves on the clinical data from Deccan hospital when the majority vote (two out of three radiologists) is used as ground truth. VGG16+ResNet50 has a slight edge in performance. Finally, in Table \ref{table:rocs3}, we have listed the area under curve for only the images with triple consensus. Again the VGG16+ResNet50 architecture has a slight edge with an area under ROC of 0.963 and area under PR curve of 0.967.

As expected, when only triple consensus images are used to test the network, the model performs better. One can argue that the images where one radiologist has disagreed with the other two are probably the more difficult cases. In other words, the ground truth itself is somewhat questionable and dropping these cases seems like a reasonable choice.

Figure \ref{fig:rocs_all} shows the ROC curves obtained using the VGG16+Resent50 pyramid network on the public development test set, clinical test with two out of three majority, and clinical test with triple consensus ground truth. Note that we are detecting normalcy, therefore the vertical axis of the ROC curve is sensitivity defined for normalcy. One key operational point identified in Fig. \ref{fig:rocs_all}(c) helps us illustrate the value of the developed network in clinical practice. The curve shows that the network can detect up to 33\% of the normal cases without misclassifying any disease cases.

Another noteworthy point is that the area under ROC curve on test set of the development data (Fig. \ref{fig:rocs_all}(a)) is smaller than that on the clinical data (Fig. \ref{fig:rocs_all}(b) and (c)). Several factors could play a role here. The size of the MIMIC-NIH test set is more than ten times the Deccan set. It is possible that a higher degree of variability exists in this larger set. Also, the ground truth here is based on NLP analysis of text report as opposed to the radiologists' consensus.

\begin{table}[t]
\caption{Area under ROC curve, and area under PR curve, on clinical data with two out of three majority ground truth (1749 images).}
\begin{tabular}{ c | c | c }
Architecture & AUC: ROC & AUC: PR  \\
\hline
VGG16+DenseNet121 pyramid & 0.916 & 0.920 \\
VGG16+ResNet50 pyramid  & 0.920 & 0.924
\end{tabular}
\label{table:rocs2}
\end{table}

\begin{table}[t]
\caption{Area under ROC curve, and area under PR curve, on clinical data with triple consensus ground truth (1271 images).}
\begin{tabular}{ c | c | c }
Architecture & AUC: ROC & AUC: PR  \\
\hline
VGG16+DenseNet121 pyramid & 0.959 & 0.965 \\
VGG16+ResNet50 pyramid & 0.963 & 0.967
\end{tabular}
\label{table:rocs3}
\end{table}

\section{CONCLUSION}

This work reports one of the first experiences with deployment of a deep learning model ``in the wild'' to detect normalcy, as defined against a comprehensive list of possible clinical findings in CXR images. The model is developed on public datasets but shown to perform accurately in the clinical settings. The ground truth for testing is completely independent from the NLP used for labeling training data and is obtained by triple consensus of radiologists. The more difficult task ahead is to develop a classifier that not only isolates the normal cases, but also pinpoints the type of finding or disease in abnormal cases, based on a comprehensive catalogue of findings.


\bibliographystyle{IEEEbib_init}
\bibliography{Ref}

\end{document}